# Generating Hierarchically Modular Networks via Link Switching

Susan Khor

## ABSTRACT

This paper introduces a method to generate hierarchically modular networks with prescribed node degree list by link switching. Unlike many existing network generating models, our method does not use link probabilities to achieve modularity. Instead, it utilizes a user-specified topology to determine *relatedness* between pairs of nodes in terms of *edge distances* and links are switched to increase edge distances. To measure the modular-ness of a network as a *whole*, a new metric called $Q_2$ is proposed. Comparisons are made between the $Q$ [15] and $Q_2$ measures. We also comment on the effect of our modularization method on other network characteristics such as clustering, hierarchy, average path length, small-worldness, degree correlation and centrality. An application of this method is reported elsewhere [12]. Briefly, the generated networks are used as test problems to explore the effect of modularity and degree distribution on evolutionary search algorithms.

## 1. INTRODUCTION

Many real-world networks of natural and man-made phenomena exhibit topological properties atypical of classical random graphs [1, 14]. Real-world networks tend to exhibit heterogeneous connectivity of the kind that, allowing for finite sizes of the networks, exhibits a power-law decay, i.e. $P(k) \sim k^{-\gamma}$, $\gamma > 1$ where $P(k)$ is the probability that a randomly selected node is linked to $k$ other nodes and $\gamma$ is the degree exponent or scaling factor. Typically $2 \leq \gamma \leq 3$. Real-world networks also tend to be more modular and have higher levels of network clustering than expected in comparable random networks. A *modular network* has identifiable subsets of nodes with a higher density of links amongst nodes within a subset than between nodes of different subsets [9, 15, 19]. *Modularity* (sometimes called *hierarchical clustering* or *community structure*) is a common characteristic of biological [7, 18] and other real-world networks [9, 22]. Network clustering refers to the cliquishness of a network and its prevalence in a network indicates dependency between links.

Hierarchical organization was proposed as the key to combine a heavy-tailed right-skewed degree distribution with high network clustering within a single network [19]. A *hierarchically modular network* is one where the nodes can be recursively subdivided into modules (subsets of unexpectedly densely linked nodes) over several scales until some atomic level is reached. An artificial instance is the *hierarchical network* model [19]. The modules in a hierarchy need not be isolated from each other, but can be interrelated subsystems of a larger encompassing whole. The "relations that hold among its parts" are, as Simon (1969) [21] emphasized, important. Hierarchical organization was also suggested as one of the pillars of the architecture of complex systems [21]. It is not surprising then that hierarchical organization is detected in many real-world networks [5].

In order to study real-world networks, it is therefore handy to have a method to generate networks with similar characteristics as real-world networks, i.e. to generate hierarchically modular networks. Like the *hierarchical random graph* model [5], our method to generate hierarchically modular networks uses a pre-specified topology to outline the hierarchical structure and guide the formation of modules. However, unlike other network generating models that employ link probabilities, whether derived internally from the network itself e.g. as a function of node degree [4], or defined externally e.g. [5], or a combination of both, e.g. [11]; our method does not deal with link probabilities directly. Instead, it utilizes the topology to determine *relatedness* between pairs of nodes in terms of *edge distances*. Nodes that belong to the same module are more related to one another than nodes that belong to different modules. Larger edge distance values are associated with more related nodes. The modularization algorithm performs link switching that favours links between more closely related nodes (according to the topology used) over less closely related nodes. Hogg (1996) [11] used an ultrametric distance (similar to our notion of edge distance) in his method to create *clustered graphs*. To measure the modular-ness of a network as a *whole*, a new metric called $Q_2$ is introduced.

## 2. THE METHOD

The networks generated by this algorithm are simple graphs, i.e. unweighted, undirected, and have no loops (self-edges) and no multiple edges.

### Algorithm

1. Create a random graph with N nodes and a given *node degree list (ndl)*. Let the resultant graph be $G_0$.

2. Randomize $G_0$ by exchanging or *switching* pairs of edges selected uniformly at random. Let the resultant graph be $G_r$.

3. Generate $T$, the *decomposition topology*.

4. Measure $aed(G_r)$, the *average edge distance* for $G_r$ relative to $T$.

5. Modularize $G_r$ by selectively exchanging pairs of edges preferentially selected at random. Edge switching is biased towards increasing edge distances relative to $T$. Let the resultant graph be $G_m$.

6. Measure $aed(G_m)$, the average edge distance for $G_m$ relative to $T$. $aed(G_m) > aed(G_r)$ is interpreted as $G_m$ being more modular than $G_r$. Using $aed(G_m)$ and $aed(G_r)$, measure modular-ness of $G_m$ with the $Q_2$ metric as follows:

$$1.0 - \frac{aed(G_r)}{aed(G_m)}.$$

**Step 1: Node degree lists and random graph $G_0$ creation.**

The degree of a node $deg(n)$ is the number of edges adjacent to $n$. A node degree list ($ndl$) enumerates the degree for all nodes in an undirected graph in ascending node label order (assumes all nodes are uniquely labeled and node $i$ represents problem variable $i$). It is common in the complex networks literature to speak of generating random graphs with a given *degree sequence*. A degree sequence of an undirected graph is its node degrees listed in non-increasing order. However, this ordering is not necessary here.

Two basic conditions for a well-formed node degree list are: (i) it must sum to an even number, and (ii) all its elements must be positive integers. The values of an $ndl$ are in the order generated by the random number generator and satisfy an additional condition: (iii) all its elements must be much smaller than the number of nodes N, and at least as large as the minimum node degree $deg_{min}$.

$G_0$ is produced as follows:

(i) A randomized (shuffled) list *all_nodes*, is made of all nodes according to their node degrees, e.g. if $deg(n_0) = 2$, then $n_0$ will appear twice in this list.

(ii) A node $x$, is chosen uniformly at random from *all_nodes* as the source of a link $e$.

(iii) Nodes other than $x$ in *all_nodes* are uniquely inspected in order of their position in *all_nodes* starting from a node, $y$, chosen uniformly at random, to find the target of $e$. A link between two nodes can be made only if the nodes are distinct from each other, the nodes are not already linked to one another and the link does not change the given node degree list. If link $e$ is made between $x$ and $y$, the frequencies of nodes $x$ and $y$ in *all_nodes* are reduced by one each.

(iv) If the target of $e$ is not found as yet, a link $\ell$ is chosen uniformly at random from the existing set of edges. Let $\ell$ connect nodes $u$ and $v$, i.e. $\ell = (u, v)$. Since loops are prohibited, $u \neq v$. Node $u$ ($v$) is made the target for $e$ provided $x \neq u$ ($x \neq v$), and a link does not already exist between $x$ and $u$ ($v$). If $u$ ($v$) is made the target of $e$, link $\ell$ is removed from the existing set of edges, and the frequency of node $v$ ($u$) in *all_nodes* is increased by one while the frequency of node $x$ in *all_nodes* is decreased by one. Nonetheless, at this point, the target of $e$ may still not be found and the search is abandoned. The random graph creation algorithm loops back to (ii) to try with another source node.

The random graph creation algorithm iterates through steps (ii) to (iv) $N^2$ times, or until *all_nodes* is emptied, i.e. a random graph with the given node degree list has been formed. If *all_nodes* is not empty after $N^2$ iterations, the node degree list is abandoned. The set of simple graph configurations for certain node degree lists can be very small, and the random graph creation algorithm above may not be successful.

**Step 2: Randomization of $G_0$.**

Using a common procedure in random network formation, pairs of links in a $G_0$ are chosen uniformly at random and exchanged if permissible, to reduce any bias inadvertently introduced into $G_0$ in step 1. Edges are exchanged only if the switch does not introduce loops or multiple-edges. Suppose the pair of original links to be switched is $(p, q)$ and $(r, s)$. Two patterns of exchange are used: $(p, s)$ and $(q, r)$; and $(p, r)$ and $(q, s)$. The given node degree list is preserved by this step.

**Step 3: Decomposition topology $T$ creation.**

The decomposition topology $T$ is used in step 5 to guide the formation of modules in a network. Previously, Clauset et al. [5] proposed the creation of *hierarchical random graphs* (random graphs with hierarchical structure) from a pre-specified topology in the form of a *dendrogram D* (a special kind of binary tree), and a set of probabilities $\{p_r\}$. The leaf nodes of $D$ represent graph nodes, while the non-leaf or internal nodes of $D$ identify groups of related graph nodes, i.e. modules. Each internal node $r$ in $D$ is associated with a probability value $p_r$. The probability of connecting two graph nodes $i$ and $j$ is $p_r$ where $r$ is the lowest ancestor node in $D$ common to $i$ and $j$. The values $p_r$ can be adjusted to favour different types of connections, e.g. short-range connections between closely related nodes versus long-range connections between distantly related nodes, or vice versa.

We use conventional binary trees, and do not require a set of pre-defined probabilities $\{p_r\}$. The function of $\{p_r\}$ is taken up by edge distance (defined in step 4) and the edge switching condition in step 5 which favours short-range connections over long-range ones where possible (although the reverse or other edge switching condition may be specified). Like the internal nodes of $D$, nodes in $T$ help identify modules – these modules are theoretical because it remains to be seen whether the actual network, $G_m$, respects them. $T$ carves out modules so that smaller modules are nested within larger modules to form a hierarchy, and there is no overlapping of territory between modules at the same level.

Figure 1 depicts $T$ for N = 20 and $ts$ = 4 (minimum size of a module). Nodes of $T$ are denoted *internal nodes*. Edges between internal nodes are *internal edges*, and a path comprised exclusively of internal edges is an *internal path*. All internal paths originate at the root of $T$. Other kinds of trees or structures could be generated for $T$.

**Step 4: Measuring edge distance $ed$, relative to $T$.**

Edge distance is a measure of the relatedness between a node pair. The distance of an edge $e = (x, y)$ is the length of the longest internal path shared by $x$ and $y$. As in a hierarchical random graph where more closely related nodes have lowest common ancestors which are situated lower in $D$ than distantly related nodes, nodes incident on edges with larger edge distance values are more related to one another in the sense that they are more likely to belong to the same module according to $T$.

The following example is with reference to Figure 1. Let $e_1 = (1, 4)$, $e_2 = (5, 4)$ and $e_3 = (4, 17)$. The longest internal paths (defined in step 3) for nodes 1, 4, 5 and 17 respectively are: ⟨10i, 5i, 2i⟩, ⟨10i, 5i, 2i⟩, ⟨10i, 5i, 7i⟩ and ⟨10i, 15i, 17i⟩. Since the longest internal paths for nodes 1 and 4 have two internal edges in common, i.e. (10i, 5i) and (5i, 2i); the edge distance of $e_1$, $ed(e_1)$, is 2. Similarly, $ed(e_2) = 1$, and $ed(e_3) = 0$. Thus, of nodes 1, 5 and 17, node 4 is more related to node 1 than to node 5, and least related to node 17. The lowest common ancestor for nodes 1 and 4

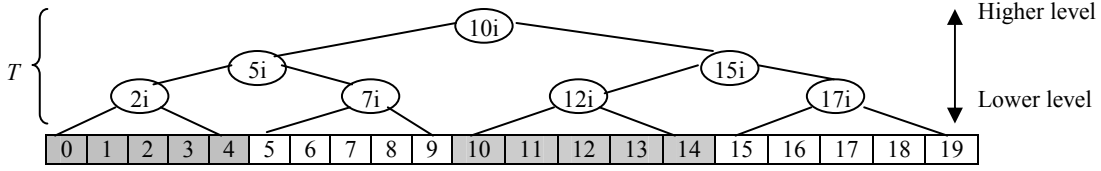

**Figure 1. A decomposition topology** $T$ for N=20 and $ts$ = 4. The internal nodes, i.e. nodes of $T$, are labeled $xi$ to distinguish them from the actual network nodes which are arranged in a row in ascending node label order at the bottom of $T$. Internal nodes identify modules. E.g. internal node 5i identifies the module encompassing nodes 0 to 9, while internal node 2i marks that nodes 0 to 4 belong to a module. The module identified by 2i is organized in $T$ to nest directly within the module identified by 5i.

is 2i, which is lower in $T$ than 5i, the lowest common ancestor for nodes 4 and 5.

The average edge distance for a graph $G$, $aed(G)$, is the sum of all edge distances divided by the number of edges in $G$:
$\frac{1}{M}\sum_{i=0}^{M-1} ed(e_i) w_i$. M is the number of links, and since networks are unweighted, $w_i$ = 1.0 for all $i$.

**Step 5: Modularization of $G_r$.**

The link switching method in step 2, with additional conditions, is used to modularize $G_r$ as follows:

(i) The *complementary edge distance* (*ced*), for all edges is calculated. The complementary edge distance for an edge $e$ is the longest internal path in $T$ $ed_{max}$, less the edge distance of $e$ plus 1, i.e. $ced(e) = ed_{max} - ed(e) + 1$. The "plus 1" ensures that all edges are included at least once in *all_edges*. A randomized (shuffled) list *all_edges*, is made of all edges according to their *ced* values, i.e. if $ced(e_0) = 2$, then $e_0$ will appear twice in this list.

(ii) Distinct pairs of edges are selected uniformly at random from *all_edges* for exchange in the step (iii). In this way, edges with smaller edge distances linking less related nodes relative to $T$ is preferentially selected for modularization.

(iii) Let $e_1 = (p, q)$ and $e_2 = (r, s)$ be the distinct pair of edges selected in step (ii). Then, the two pairs of alternative edges are $e_3 = (p, r)$ and $e_4 = (q, s)$, and $e_5 = (p, s)$ and $e_6 = (q, r)$. Let the edge distance for edge $e_i$ be $ed_i$. If an edge is a loop or introduces a multiple-edge, its edge distance is -1. Let $ped_{ij}$ be the product of edge distances of a pair of edges $i$ and $j$, i.e. $ped_{ij} = ed_i \times ed_j$. The original edge pair ($e_1$, $e_2$) is exchanged with the edge pair that has the larger $ped_{ij}$ value larger than $ped_{12}$. Thus, ($e_1$, $e_2$) is switched to ($e_3$, $e_4$) only if $ped_{34} > ped_{12}$ and $ped_{34} \geq ped_{56}$, and ($e_1$, $e_2$) is switched to ($e_5$, $e_5$) only if $ped_{56} > ped_{12}$ and $ped_{56} \geq ped_{34}$.

(iv) If a switch is made in step (iii), the modularization algorithm goes to step (i), otherwise it loops to step (ii).

The modularization algorithm iterations through steps (i) to (iv) $P_g \times [M + N(N-1)/2]$ times. $P_g$ is a parameter for modularizing networks. The number of edges M, is included in the number of times the modularization algorithm is iterated to accommodate graphs with the same number of nodes N, but significantly different number of edges. Increasing the number of times the modularization algorithm is applied on a network need not result in a more modular network because a network's degree distribution in part constraints a network's structural possibilities. For example, node degree lists 13 and 14 (section 3) have the two largest M amongst all the *ndl*s (Table 1), but their $Q$ and $Q_2$ (Figure 3) values are on the smaller side. This is expectable since nodes with unusually high degree have no choice but to make inter-module links.

The modularization algorithm has the effect of reducing the number of edges with smaller edge distances, and increasing the number of edges with larger edge distances (Figure 2). However, step (iii) does not necessarily favour the preservation of edges with larger edge distances over edges with smaller edge distances, and permits for instance the exchange of a pair of edges with edge distances 1 and 8 with a pair of edges with edge distances 4 and 5. This exchange favours the formation of larger modules over the formation of smaller modules.

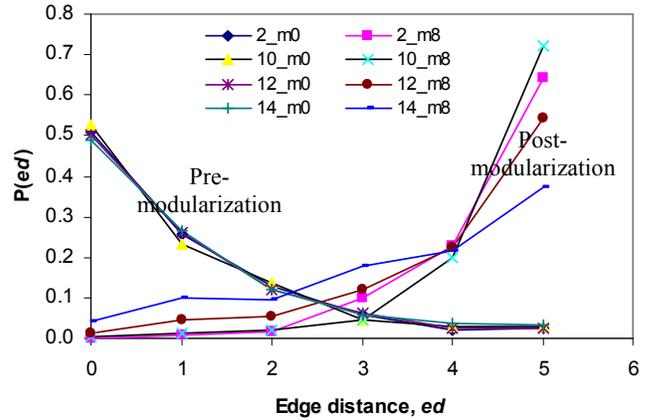

**Figure 2. Change in edge distance distribution due to modularization.**

**Step 6: Measuring modular-ness with $Q_2$.**

It is proposed here that modular-ness of a network $G_B$ be assessed relative to the modular-ness of a comparable network $G_A$ by $Q_2$ as follows: $1.0 - \frac{aed(G_A)}{aed(G_B)}$. $Q_2$ is 0.0 if $G_B$ is as (not) modular as $G_A$, i.e. $aed(G_B) = aed(G_A)$, $Q_2$ is > 0.0 if $G_B$ is more modular than $G_A$, i.e. $aed(G_B) > aed(G_A)$, and $Q_2$ is < 0.0 if $G_B$ is less modular than $G_A$, i.e. $aed(G_B) < aed(G_A)$. In this paper, $G_r$ is $G_A$, and $G_m$ is $G_B$. In its usage here, $Q_2$ compares what is presumed to be a modularized graph with its previous random and therefore likely less modular version. It is possible, as in a fully connected simple graph, that there are no alternative simple graph configurations. In this case, $Q_2$ will be 0.0, which is appropriate since all nodes in a fully connected graph belong to the one same module and thus does not exhibit modularity.

## 3. AN EXAMPLE

To illustrate, the method in section 2 is applied to eight node degree lists (*ndl*s). For each network, the number of nodes N is 200, and the number of links M is given in Table 1. Minimum node degree $deg_{min}$ is set to three to induce the formation of connected graphs so that all nodes of a network belong to a single graph component.

The *ndl*s were produced by rounding the values generated by the *randht.m* procedure (version 1.0.2) provided online by A. Clauset. Two different kinds of distributions are used: normal (*ndl*s 1 and 2) and power-law with degree exponents 4.0 (*ndl*s 9 and 10), 3.0 (*ndl*s 11 and 12) and 2.6 (*ndl*s 13 and 14). The characteristics of the *ndl*s are summarized in Table 1. The degree distribution of *ndl*s 1 and 2 have little to no skew, mean = median = mode. The degree distribution curves of the other *ndl*s are right-skewed, mean ≥ median ≥ mode. The standard deviations noticeably increase going down the list of *ndl*s in Table 1.

**Table 1. Node degree list summary statistics, N=200**

| ndl | Min | Max | Mean | Std. dev | Mod | Median | M |
|---|---|---|---|---|---|---|---|
| 1 | 3 | 9 | 6.05 | 1.0786 | 6 | 6 | 605 |
| 2 | 3 | 9 | 6.04 | 1.0459 | 6 | 6 | 604 |
| 9 | 3 | 19 | 4.60 | 2.2573 | 3 | 4 | 460 |
| 10 | 3 | 15 | 4.36 | 1.8595 | 3 | 4 | 436 |
| 11 | 3 | 36 | 5.74 | 4.3006 | 4 | 4 | 574 |
| 12 | 3 | 53 | 5.25 | 4.3822 | 3 | 4 | 525 |
| 13 | 3 | 50 | 6.94 | 7.0673 | 3 | 5 | 694 |
| 14 | 3 | 68 | 6.74 | 7.6265 | 4 | 5 | 674 |

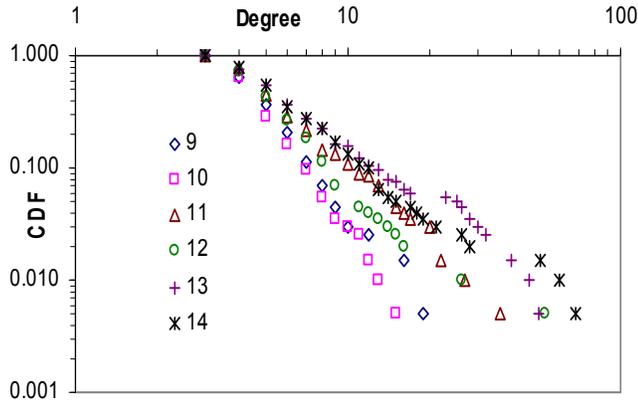

**Figure 3. The complementary cumulative distributions (CDF) of *ndl*s 9 to 14 on a log-log plot.**

Figure 3 depicts the complementary cumulative distribution function (CDF): $P(x) = Pr(X \geq x)$ of *ndl*s 9 to 14 on a double log-scale. Whether these distributions are best characterized by power-laws is less important than that they are heavy-tailed and right-skewed. The doubly logarithmic scale is used as it is a convenient form to depict the distributions. Power-law identification for the *ndl*s is hampered by the fact that N=200 which is small compared to the hundreds of nodes in real-world scale-free networks. $deg_{min}$ which is 3, is also smaller than 5 which increases the error associated with our method of generating the *ndl*s [6]. Nevertheless, these networks are generated for the purpose in [12] and using a larger N would substantially increase simulation time without necessarily producing more relevant insight. The *ndl*s may be produced by network growing models such as the many variations of the preferential attachment model [2], other generalized random graph models [13], and using biologically inspired mechanisms such as duplication and divergence. This is a possible next step.

Step 1 successfully constructed $G_0$ for every *ndl*, in part because $deg_{min}$ is three, and $P_e$, the fraction of actual to all possible links, which ranges from 0.022 to 0.035, is very low.

Attempts at edge switching in step 2 are made $0.125 \times N(N-1)/2$ times, which is 2,487 for networks in this paper. The most number of links a network in Table 1 has is 694, so each link would have had a chance to switch.

All networks in this paper use the same $T$ which is created as follows in step 3: nodes of $G_0$ are arranged by node label in a string, and this string is recursively split into two (almost) equal sized halves at node labeled $x$ until the remaining portion is smaller than some *size ts*, which is set to four here. The set of all $x$ node labels derived from this process forms $T$.

In step 5, $P_g$ is set to 0.8.

## 4. MODULARITY

To verify that the modularization algorithm in step 5 does indeed modularize a network, $Q$ values for the first three highest levels for the networks generated from the *ndl*s in section 3 are taken before and after step 5. The $Q$ metric was introduced in [15] as a measure of modularity given a certain division of a network. $Q = \frac{s^T B s}{2m}$; where $s$ is a column vector of ±1 elements representing a particular division of a network into two candidate modules, $s^T$ is the transpose of $s$, and $B$ is a real symmetric matrix called the *modularity matrix* with elements $B_{ij} = A_{ij} - \frac{k_i k_j}{2m}$. $A$ is the adjacency matrix for the network where $A_{ij} = e$ means $e$ links exists between nodes $i$ and $j$, $k_i$ and $k_j$ are the respective degrees of nodes $i$ and $j$, and $m$ is the total number of links in the network. Elements of $B$ reflect the statistical surprising-ness of links relative to what could be expected by random chance. A positive (negative) $Q$ value indicates that a network has fewer (more) links than expected between its two divisions as delineated by $s$. Each row and each column of $B$ sum to 0 which assures the existence of an all ones eigenvector with an eigenvalue of zero. A network is indivisible when no other $s$ but the all ones vector produces a non-negative $Q$ value. Indivisible networks have a $Q$ value of 0.0. Optimally divided networks have a $Q$ value of 1.0.

Since $T$s in this paper are binary trees which subdivide modules into two more or less equal halves, the $s$ vectors for $Q$ follow suit. For example, to calculate the highest level $Q$ value for a network with $T$ in Figure 1, $s$ has 20 elements with +1 in its top half and -1 in its bottom half. There are two $Q$ values at the second highest level. The $Q$ value for one module is calculated for nodes 0 to 9, and its $s$ has 10 elements with +1 in its top half and -1 in its bottom half. The $Q$ value for the other module is similarly calculated for nodes 10 to 19.

Table 2 gives a sample of the $Q$ and $Q_2$ values before modularization (m0) and after modularization (m8) for networks generated from four different *ndl*s. The degree distribution curve for *ndl* 2 is bell-shaped while it is right-skewed and heavy-tailed for the other three *ndl*s (section 3). Prior to modularization, the $Q$ values are negative and close to 0.0000 and the $Q_2$ values by definition is 0.0000. After the modularization algorithm in step 5 is applied with $P_g = 0.8$, the $Q$ values for the first three highest levels increase significantly towards 1.0. The average edge distance (*aed*) values for modularized networks are at least 3.5 times that of non-modularized networks. As such, $Q_2$ values of the modularized networks rise significantly above 0.0000 (Figure 4). In short, the networks do become more modular after step 5, and the $Q_2$ measure does indicate increase in modular-ness of a network.

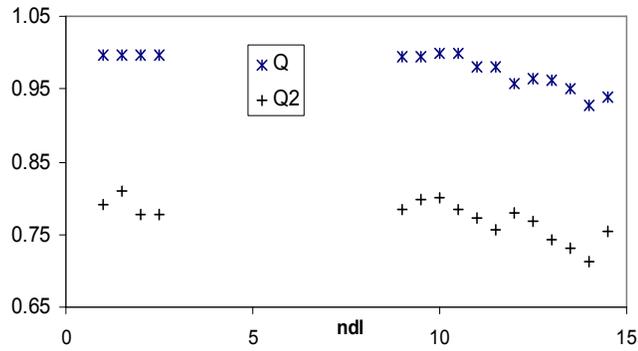

**Figure 4. $Q$ and $Q_2$ values for modularized networks.**

However, while the correlation between the $Q$ values for the highest level and corresponding $Q_2$ values is strong (0.8487 for the networks in section 3), the two measures need not necessarily rank networks by modular-ness in the same order. This reflects the semantic difference between $Q$ and $Q_2$. While $Q$ measures modular-ness of a network with respect to a particular division of the network at a single level, $Q_2$ considers the modular-ness of a network in its entirety with respect to a particular decomposition topology $T$.

N1 = {(0, 1), (0, 2), (0, 3), (0, 4), (4, 6), (4, 5), (4, 7)}, and N2 = {(0, 1), (0, 2), (0, 4), (2, 3), (4, 5), (4, 6), (6, 7)}, where $(u, v)$ represents a link between nodes $u$ and $v$, are two networks of the same size (N=8, M=7). Assuming $T$ is a perfect binary tree with $ts$ = 2, N1 and N2 have the same $Q$ value (0.7143) for the highest level, but compared with the same random graph N1's $Q_2$ value is smaller than N2's rightly reflecting the fact that N2 is more modular than N1 relative to $T$. Hence, the $Q$ and $Q_2$ metrics are distinguishable from each other.

## 5. HIERARCHY

A *clustering coefficient spectrum* that is inversely related with node degree is interpreted as indicative of hierarchical organization [19]. A network's clustering coefficient spectrum $C(k)$ values are obtained by averaging the clustering coefficient of node $i$ $C_i$ for all $i$ with degree $k$. $C_i$ is the ratio of actual to possible links amongst a set of nodes: $C_i = \dfrac{2E_i}{k_i(k_i - 1)}$. $E_i$ is the number of links between node $i$'s $k$ neighbors, and $k(k-1)/2$ is the number of possible (undirected single) links between $k$ nodes [26]. Figure 5 shows that modularization significantly increases the network clustering coefficient $C$ for all networks regardless of degree distribution type.

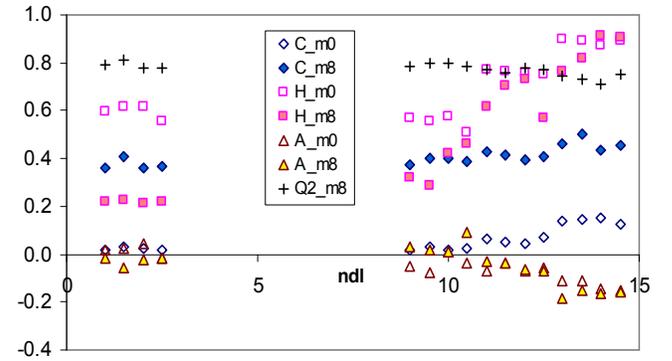

**Figure 5. Network characteristics for non-modularized (_m0) and modularized (_m8) networks.**

However after modularization, $C(k)$ values for broad connectivity networks (*ndl*s 9 to 14) become significantly more inversely related with $k$, while $C(k)$ values for random connectivity networks (*ndl*s 1 and 2) show almost uniform increases independent of $k$ (Figure 6). This implies that after modularization step following the binary tree decomposition topology $T$, networks with normal degree distribution (*ndl*s 1 and 2) are less hierarchically organized than networks with right-skewed degree distribution (*ndl*s 9 to 14). This observation is supported by the $H$ metric [22] which decreases more sharply for *ndl*s 1 and 2 than for the other *ndl*s (Figure 5). The $H$ metric identifies hierarchical architecture in a network through the concept of *hierarchical path* [10]. "A path between nodes $a$ and $b$ is hierarchical if (1) the degrees of vertices along this path vary monotonously from one vertex to the other, or

**Table 2. Network modular-ness pre- (m0) and post- (m8) modularization.**

| N = 200 | $Q$ values for the first three highest levels | | | | | | | *aed* | $Q_2$ |
|---|---|---|---|---|---|---|---|---|---|
| 2_m0  | -0.0232 | -0.0069  | -0.1067  | -0.1940 | -0.0900 | -0.2407 | 0.0220  | 0.8940 | 0.0000 |
| 2_m8  | 0.9966  | 0.9666   | 0.9934   | 0.9725  | 0.9690  | 0.9339  | 0.9731  | 4.4669 | 0.7999 |
| 10_m0 | -0.0554 | 0.0893   | -0.0750  | -0.5936 | 0.1023  | -0.1478 | -0.4025 | 0.9014 | 0.0000 |
| 10_m8 | 0.9905  | 0.9715   | 0.9662   | 0.9268  | 0.9611  | 0.9456  | 0.9799  | 4.5757 | 0.8030 |
| 12_m0 | -0.0097 | -0.11097 | -0.08526 | -0.5147 | -0.1736 | 0.0636  | -0.0494 | 0.9219 | 0.0000 |
| 12_m8 | 0.9732  | 0.8371   | 0.9555   | 0.6585  | 0.9636  | 0.9636  | 0.9518  | 4.1295 | 0.7768 |
| 14_m0 | 0.0200  | 0.0394   | -0.1286  | 0.2849  | -1.0149 | -0.8272 | -0.1720 | 0.9896 | 0.0000 |
| 14_m8 | 0.9162  | 0.8118   | 0.7638   | 0.8887  | 0.6762  | 0.7780  | 0.5917  | 3.5460 | 0.7209 |

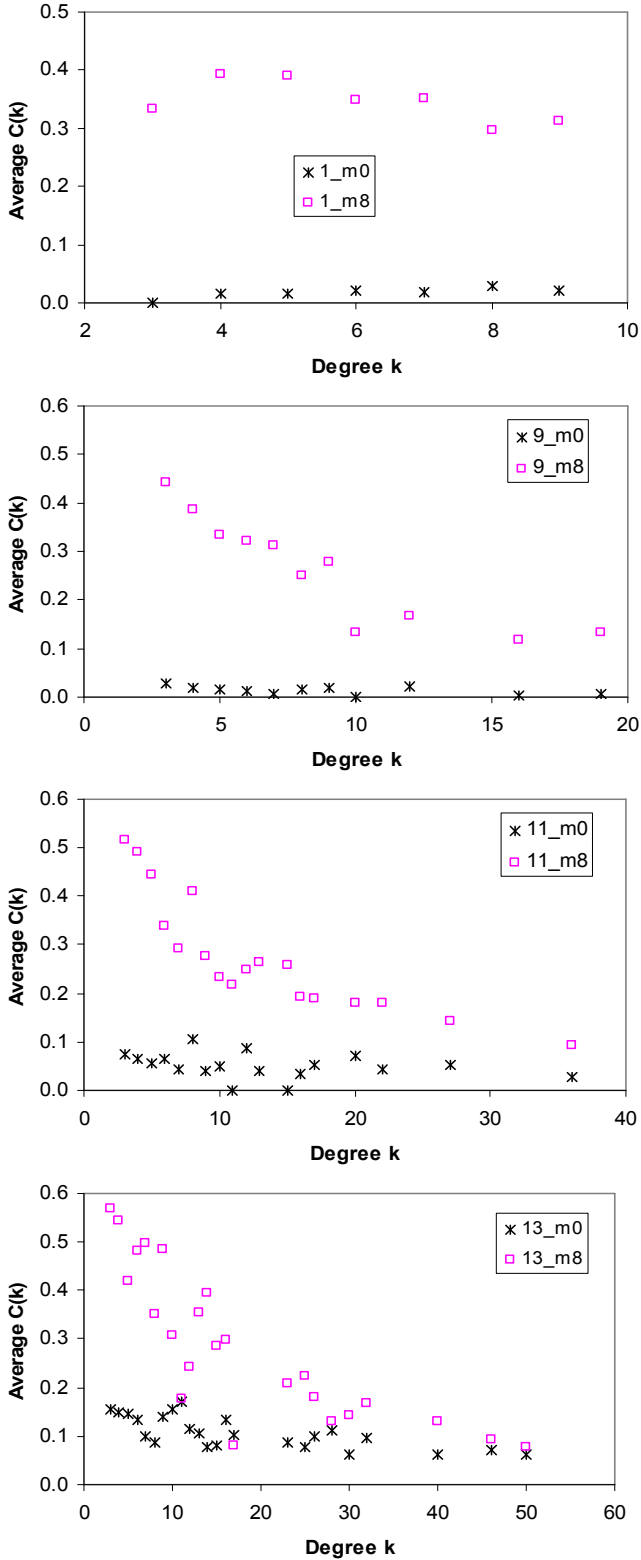

**Figure 6. Clustering coefficient spectrum for non-modularized (_m0) and modularized (_m8) networks. The network with normal degree distribution (1_) is less hierarchically organized than the other networks. Note the difference in the x-axis.**

(2) vertex degrees first monotonously grow, reach maximum value, and then monotonously decrease along the path. Let the fraction *H* of the shortest paths in a network be hierarchical. Then this number *H* can be used as a metric of a hierarchical topology [27]. If *H* of a network is sufficiently close to 1, the network has a pronounced hierarchical organization." [8].

Thinking in terms of hierarchical paths can help to explain the lack of hierarchical organization in networks with normal degree distribution (*ndl*s 1 and 2). These networks lack hubs, nodes with unusually high degree, through which paths between pairs of nodes especially distant pairs of nodes (in terms of their label differences, e.g. the leftmost and the rightmost nodes in Figure 1), can route through. Therefore, they have fewer hierarchical paths and lower *H* values. Post-modularization, the *H* values (Figure 5) become positively correlated with Max node degree (Table 2).

## 6. SMALL-WORLDNESS

The distance between two nodes is the number of edges or length of a shortest (geodesic) path connecting them. Average path length is the average distance between all node pairs in a network. The diameter (maximum) and average path lengths (*apl*) for networks generated from *ndl*s 1, 2, 9 and 10 increased significantly after modularization (Figure 7). This can be explained by the lack of hubs, or nodes with large enough degree values to span module boundaries and become short-cuts in a network. Interestingly, for modularized networks, there seems to be an inverse relationship between *H* and *apl*. In particular, *H* for *ndl*s 1, 2, 9 and 10 declines as their *apl* increases.

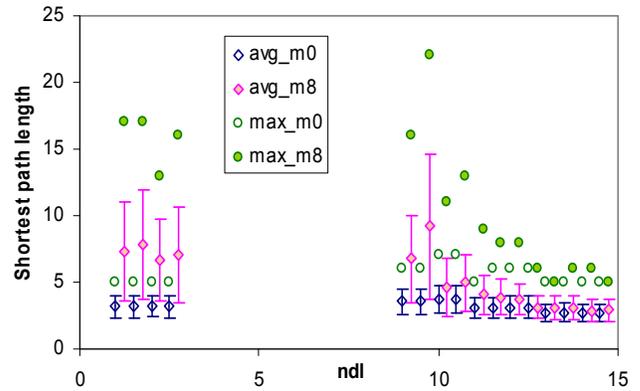

**Figure 7. Maximum and average path length for non-modularized (_m0) and modularized (_m8) networks.**

A network exhibits the *small-world* property if the network has a high clustering coefficient and if a short path connects most pairs of nodes in the network or more precisely, if the average distance (path length) between nodes, grows logarithmically or slower with network size for fixed mean degree [26]. Many artificial and natural networks exhibit the small-world property, e.g. [3, 20, 23, 24, 25].

Modularization increases network clustering for all networks (Figure 5). In this partial sense, all networks become smaller worlds compared to their pre-modularized state (when network clustering is quite low). However, due to their smaller *apl* values, modularized networks generated from *ndl*s 11 to 14 are smaller worlds compared to the other modularized networks in this paper.

## 7. DEGREE CORRELATION

Degree correlation refers to the correlation between the degrees of connected nodes in a network. Degree correlation of a network (the *A* values in Figure 5) is measured using Eq. 4 in ref. [16] which measures the correlation of degrees at either end of an edge. Networks with negative *A* values are termed *disassortative* (nodes tend to link to other nodes unlike themselves, e.g. high degree nodes are more likely to connect with low degree nodes), while networks with positive *A* values are termed *assortative* (nodes tend to link to other nodes like themselves). Collaborative and social networks tend to show assortative mixing by degree, while biological networks and the World Wide Web show disassortative mixing by degree [14].

No significant change in network degree correlation is detected between per- and post-modularization networks (Figure 5). This is not surprising since no effort is made to either increase or decrease degree correlation in a positive or negative manner (but see Appendix A for a way to vary assortativity). However, networks from *ndls* 11 to 14 are slightly more dis-assortative ($A < 0$) than the other networks.

Degree correlation of a network can also be inferred by observing how the average degree of nearest neighbours of nodes with degree $k$ ($k_{nn}(k)$) change as a function of $k$ (we follow Eq. 6 in [7]). If this function is constant, degrees of connected nodes are uncorrelated. An increasing $k_{nn}(k)$ with increasing $k$ means that on average, degree of nodes neighbouring $k$-degree nodes increase with increases in $k$, and this implies assortativity. A decreasing $k_{nn}(k)$ with increasing $k$ means on average, degree of nodes neighbouring $k$-degree nodes decrease with increases in $k$, and this implies disassortativity. By this definition, the networks generated from *ndls* 11 and 13 show slight disassortativity, while the degrees of neighbouring nodes in the other networks appear uncorrelated (Figure 8).

## 8. NODE CENTRALITY

Centrality of a node measures the number of shortest paths between other nodes that traverses the node [14]. Prior to modularization, hubs or nodes with high degree tend to occupy a central position in all networks. After modularization however, the positive correlation between node degree and *centrality* declines more substantially for networks from *ndls* 1, 2, 9 and 10 than for the other test problems (Figure 9).

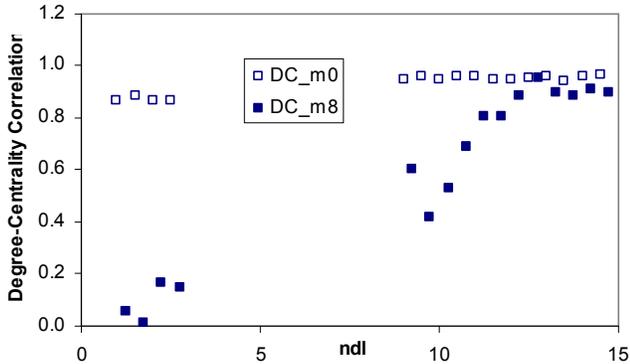

Figure 9. Correlation between node degree and centrality.

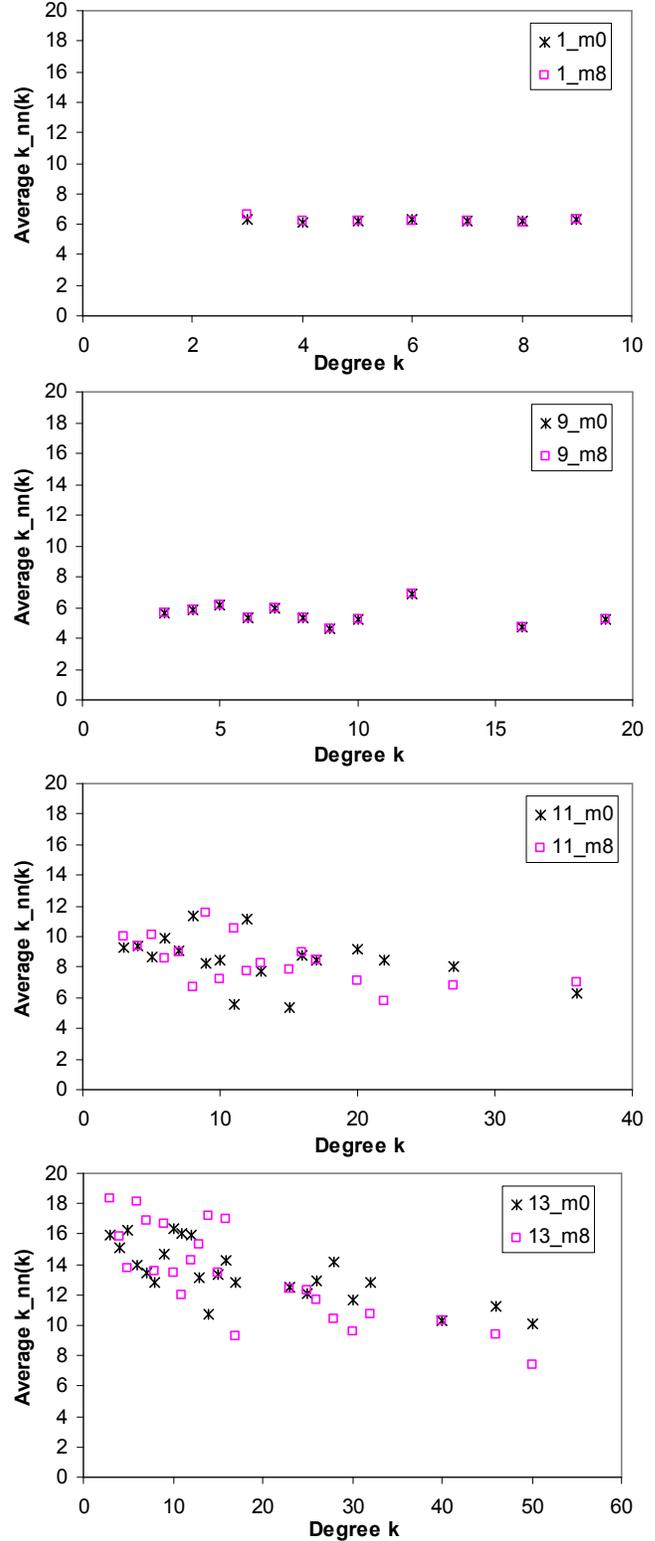

Figure 8. Nearest-neighbours spectrum for non-modularized (_m0) and modularized (_m8) networks. Note the difference in the x-axis.

## 9. DISCUSSION

We have proposed a link switching modularization method based on a pre-specified topology and edge distance (section 2) and applied the algorithm on eight different node degree lists (*ndl*) generated at random from different probability distributions (section 3). Several key network structural characteristics: modularity, hierarchy, clustering, path length, small-worldness, degree correlation and centrality were examined and the following were observed:

(i) In general, networks generated from *ndls* 11 to 14 (Group A) exhibit different structural characteristics from networks generated from *ndls* 1, 2, 9 and 10 (Group B). This suggests the influence of degree distribution on the other structural characteristics.

(ii) After modularization, all networks have higher $Q$ and $Q_2$ values (section 4). However, Group A networks show more hierarchical organization (section 5), have smaller diameters, have shorter average path lengths, are smaller-worlds (section 6), exhibit more degree correlation between neighbouring nodes (section 7), and have higher degree-centrality correlation (section 8) than do Group B networks. Coincidently, *ndls* 11 to 14 are those generated from a power-law distribution with degree exponents 2.6 and 3.0 (section 3).

An application of the link switching modularization method proposed here is reported elsewhere [12]. Briefly, the generated networks are used as test problems to explore the effect of modularity and degree distribution on evolutionary search algorithms. Further work is planned in this area of connecting problem structure, in terms of the topological characteristics of their constraint networks, and evolutionary search.

A relevant question in this respect is how well do the networks produced in this paper reflect the real-world search problems? To answer this question, further investigation of real-world search problems is first needed, and refs. [11, 25] are two promising starting points in this direction. It is also encouraging to find that networks in Group A have small-world tendencies since several benchmark search problems [25], including the protein contact map which has been used for protein folding are also found to be small-worlds [23].

However, in terms of biological networks, we detect a slight mismatch between networks in Group A and the relationship between modularity, hierarchy and degree correlation in ref. [21] which associates anti-hierarchical organization with disassortativity, and disassortativity with modularity (isolation of parts) [17]. This discrepancy may be due in part to differences in network construction, and may warrant further investigation.

scale-free networks: From random to assortative. *Physical Review E 70*, 066102.
[28] Zhou, S. and Mondragon, R.J. (2007) Structural constraints in complex networks. *New Journal of Physics 9*, 173.

## Appendix A. Varying Degree Assortativity

In this section, the algorithm described in section 2 is modified to produce networks with varying degree correlation (section 7). The basic idea behind this modification is to control links between nodes of high degree. To increase disassortativity, fewer to no links are permitted between nodes of high degree. To increase assortativity, more to all possible links are made obligatory between nodes of high degree. This modification bears resemblance to rich-club connectivity [28], and to an extension mentioned in [27] to increase clustering.

We illustrate this modification with the following three cases on test problems 11 to 14 described in section 3. All three cases involve ten nodes with the highest degrees[1]. Let this node set be R.

(i) "Top 10 0.75": Nodes in R are randomly linked with probability 0.75, and these links cannot be removed.

(ii) "Top 10 0.25": Nodes in R are randomly linked with probability 0.25, and these links cannot be removed.

(iii) "Top 10 0.00": No links are permitted between nodes in R.

Additional links between nodes in R may be created in cases (i) and (ii). The effect of this modification is compared to the "null" case, which is produced by the original algorithm, i.e. sans this modification.

As in section 7, degree correlation of a network is measured using Eq. 4 in ref. [16]. Figure A1 compares the degree correlation of networks produced under the above three conditions. Condition (iii) made the networks more disassortative than the "null" case networks. Conditions (i) and (ii) while not exhibiting a significant difference between them, nevertheless increased assortativity or decreased disassortativity of the networks. Hence, our modification produces the desired effect.

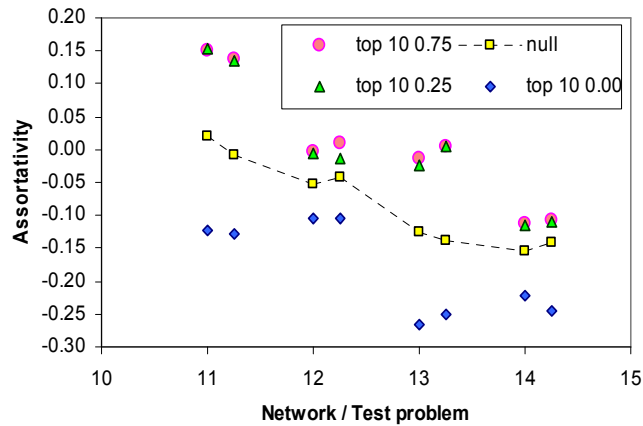

**Figure A1. Degree-degree correlation (A)**

---
[1] Nodes are sorted in descending order of their degree, and the first ten nodes in this list are made members of R.

Further, network modularity (Q2) and clustering (C) are not or only slightly affected by the change in degree correlation brought about by the modification (Figure A2). Thus, our method is able to produce modular networks with prescribed degree sequence, high clustering, and varying degree correlation.

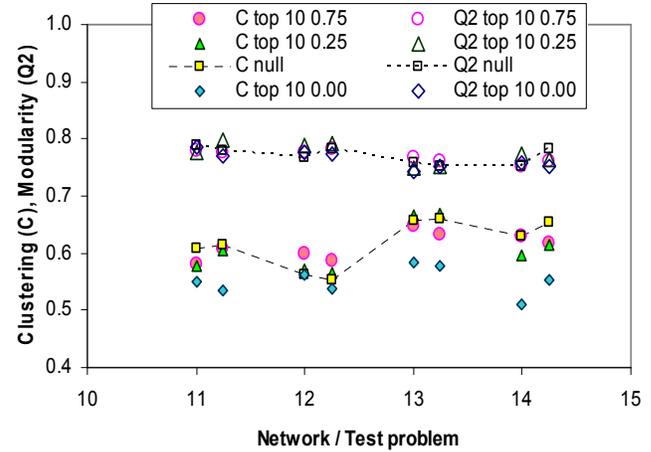

**Figure A2. Clustering (C) and Modularity (Q2)**

The networks produced with the modified algorithm also demonstrate the association made in [22] between negative degree correlation or disassortativity and anti-hierarchical organization. Networks produced under condition (iii) are more disassortative and have smaller H values than the "null" networks, while networks produced under conditions (i) and (ii) which tend towards assortativity are more hierarchical and have larger H values (Figure A3).

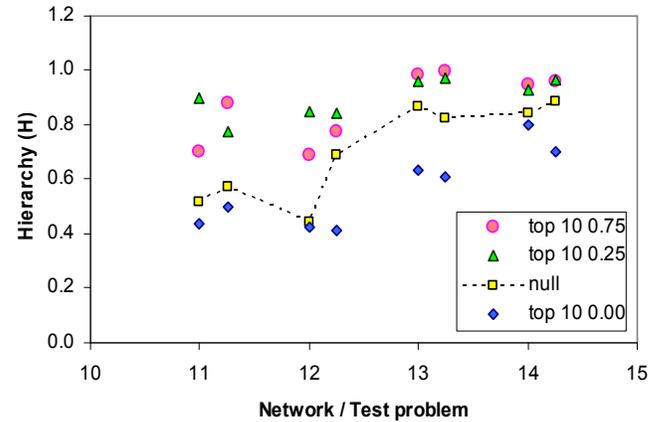

**Figure A3. Hierarchy (H)**

Figure A4 compares the median path length amongst nodes in different degree quartiles averaged over the networks. The quartiles are formed as follows: (i) unique degree values are sorted in ascending order, and (ii) this sorted list is divided into four (almost) equal parts. Quartile 1 nodes are those with degree values larger than or equal to the minimum value in the upper quartile of this sorted list (Quartile 1 nodes are those with higher degrees). Quartile 2 nodes are those with degree values larger than or equal to the median of this sorted list. Quartile 3 nodes are those with degree values larger than or equal to the minimum value of the lower quartile of this sorted list. Quartile 4 comprises all nodes in the network. We observe that the difference in average median path length (AMPL) is larger in the first two quartiles (1 and 2), than in the last two. Also, networks produced under conditions (i) and (ii) have shorter median path lengths than case (iii) networks on average. This is expected given the way the networks were constructed.

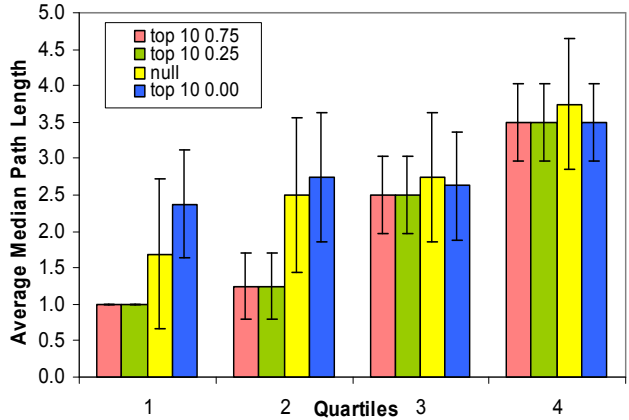

**Figure A4. Average Median Path Length by Quartile**
Error bars indicate ± one standard deviation.

Figure A5 give the diameter (longest shortest path length) of the networks and Figure A6, the average and median path length of the networks. Distinct differences between case (i) and (ii) networks and case (iii) networks in these figures are not discernable.

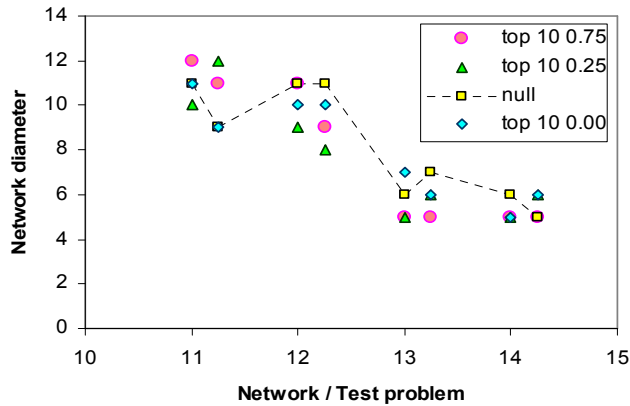

**Figure A5. Network diameter**

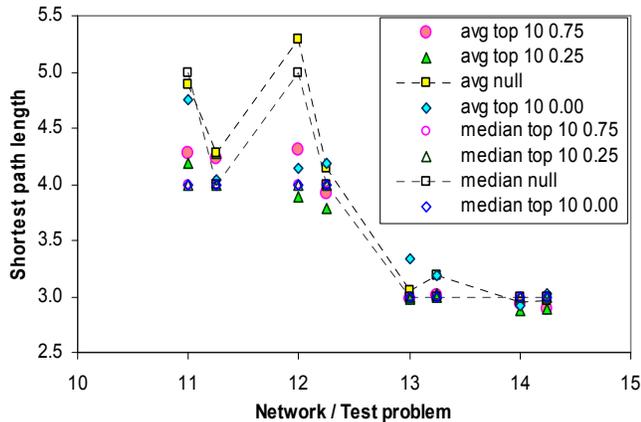

**Figure A6. Average and Median Path Length**

Finally, we confirm that it is necessary to use nodes of high degree for R to vary degree correlation. Figure A7 gives the degree correlation for networks produced under the same three conditions but with ten randomly selected R nodes. The degree correlations of these networks are indistinguishable from the "null" network.

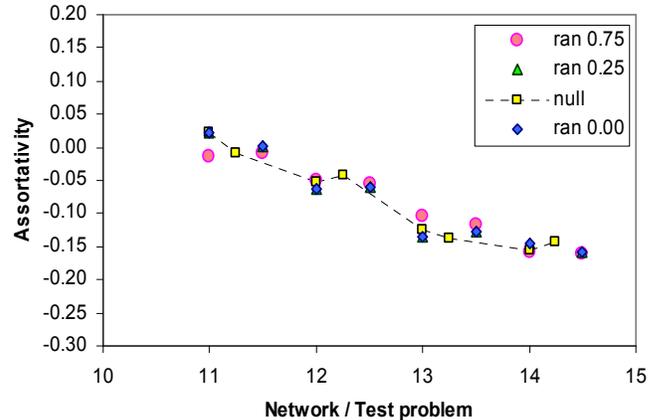

**Figure A7. Random nodes for R**

## Appendix B. Equal-only and Unequal-only problems

Using the method described in Appendix A herein, we conducted experiments that support the hypothesis that given similar conditions (i.e. number of nodes, number of links, degree distribution and clustering), fewer colors are needed to color disassortative than assortative networks[2]. This result contradicts our previous results somewhat in the sense that shorter characteristic path lengths amongst hubs were viewed favourably in [12] for quick resolutions, and leads us to the insight that *search difficulty for problems with broad degree distribution can vary considerably depending on the degree of separation between hubs, and whether the problem constraints are equal-only, e.g. the problems in [12], or unequal-only, e.g. graph-coloring.*

---

[2] This point has been developed further in relation to biological networks and ref. [17]. See author's website for further updates on this front.